**Synthesis of intrinsic magnetic topological insulator $MnBi_{2n}Te_{3n+1}$ family by chemical vapor transport method with feedback regulation**

Heng Zhang[1,2], Yiying Zhang[1,2], Yong Zhang[1,3], Bo Chen[1,2], Jingwen Guo[1,2], Yu Du[1,2], Jiajun Li[1,2], Hangkai Xie[1,2], Zhixin Zhang[1,2], Fuwei Zhou[1,2], Tianqi Wang[1,2], Wuyi Qi[1,2], Xuefeng Wang[1,3], Fucong Fei[1,4,5,*], Fengqi Song[1,2,5,*]

[1] *National Laboratory of Solid State Microstructures, Collaborative Innovation Center of Advanced Microstructures, Nanjing University, Nanjing 210093, China.*

[2] *School of Physics, Nanjing University, Nanjing 210093, China.*

[3] *School of Electronic Science and Engineering, Nanjing University, 210093 Nanjing, China.*

[4] *School of Materials Science and Intelligent Engineering, Nanjing University, Suzhou 215163, China.*

[5] *Institute of Atom Manufacturing, Nanjing University, Suzhou 215163, China.*

E-mail: feifucong@nju.edu.cn; songfengqi@nju.edu.cn (* corresponding authors)

**Abstract:**

$MnBi_{2n}Te_{3n+1}$ (MBT) is a representative family of intrinsic magnetic topological insulators, in which numerous exotic phenomena such as the quantum anomalous Hall effect are expected. The high-quality crystal growth and magnetism manipulation are the most essential processes. Here we develop a modified chemical vapor transport method using a feedback-regulated strategy, which provides the closed-loop control of growth temperature within ± 0.1 °C. Single crystals of $MnBi_2Te_4$, $MnBi_4Te_7$, and $MnBi_6Te_{10}$ are obtained under different temperature intervals respectively, and show variable tunability on magnetism by finely tuning the growth temperatures. Specifically, the cold-end temperatures not only vary the strength of antiferromagnetic coupling in $MnBi_2Te_4$, but also induce magnetic ground state transitions from antiferromagnetism to ferromagnetism in $MnBi_4Te_7$ and $MnBi_6Te_{10}$. In $MnBi_2Te_4$ with optimized magnetism, quantized transport with Chern insulator state is also realized at the low field of 3.7 T. Our results provide a systematic picture for the crystal growth and the rich magnetic tunability of MBT family, providing richer platforms for the related researches combining magnetism and topological physics.

## 1. Introduction

The interplay between magnetic order and the topological electronic structure in magnetic topological insulators (MTIs) gives rise to many novel quantum phenomena, such as the quantum anomalous Hall effect (QAHE), axion insulator phase, and magnetic Weyl semimetals.[1-9] $MnBi_2Te_4$ is a respective material of intrinsic MTI holding van der Waals layered structures stacking by Te-Bi-Te-Mn-Te-Bi-Te septuple layers (SLs).[10-15] The magnetic moments in $MnBi_2Te_4$ stems from the $Mn^{2+}$ ions, which shows interlayer antiferromagnetic (AFM) exchange interactions between adjacent SLs and ferromagnetic (FM) exchange interactions into each SL, forming an A-type AFM magnetic ordering with the Néel temperature ($T_N$) of 24 K.[16-18] Recently, $MnBi_2Te_4$ has attracted much attention for the experimental observation of QAHE, axion insulator states, the high-Chern-number QHE without Landau levels, and anomalous Landau quantization.[11, 19-22] To tune the behavior of these exotic topological states and realize more interesting phenomena, the magnetic exchange coupling plays an important role, which is hard to adjust in $MnBi_2Te_4$ due to its strong AFM coupling. Based on $MnBi_2Te_4$, several derivative MTI materials such as $MnBi_4Te_7$ and $MnBi_6Te_{10}$ have also been discovered and widely investigated.[23-29] These two materials are identified as natural superlattice heterostructures by respectively inserting one and two $Bi_2Te_3$ quintuple layers between adjacent SLs in $MnBi_2Te_4$. The interlayer AFM coupling in $MnBi_4Te_7$ and $MnBi_6Te_{10}$ is weakened because of the widening of the SL interlayer distance.[25, 30] Currently, several growth methods have been developed to grow the $MnBi_{2n}Te_{3n+1}$ (MBT) family. The more commonly used method is the self-flux method with a stoichiometric molar ratio of elements or extra $Bi_2Te_3$ as flux.[17, 25, 31-32] However, the narrow temperature interval for single-crystal growth (around 10 °C for $MnBi_2Te_4$, 2 °C for $MnBi_4Te_7$, and narrower for $MnBi_6Te_{10}$) greatly increases the difficulty in growing high-quality crystals and the further magnetic modulation of the MBT crystals by changing synthesis parameters, which are the two essential processes for researches of MTI.[23, 30, 33-34] Till now, various magnetic properties of $MnBi_4Te_7$ and $MnBi_6Te_{10}$ have been reported in different reports. For $MnBi_4Te_7$, in addition to AFM transitions with $T_N$ around 13 K and 12 K, FM-like states below 10.5 K and 5 K are found.[23, 27, 30, 35-37] For $MnBi_6Te_{10}$, both the AFM phase with $T_N$ around 10 K and the FM phase with $T_C$ around 13 K and 12K are synthesized through the self-flux method using different growth strategies.[24, 38-41] Given the differences in sample preparation parameters among different research teams in these works, it is difficult to compare these samples horizontally and map out a systematic picture of the effect of growth parameters on the magnetic properties of the as-grown MBT crystals. In addition, the $MnBi_2Te_4$ synthesized by the chemical vapor transport (CVT) method has also been reported and further shows different magnetic properties compared with the flux-grown samples,[42-43] though the CVT growth of $MnBi_4Te_7$ or $MnBi_6Te_{10}$ phases and the systematic magnetic property manipulation among MBT family are still lacking. Therefore, one can see that the reports of various magnetic properties modulation in the MBT family are still sporadic cases rather than systematic conclusions till now, and the systematical picture for the varying of the magnetic properties through parameter controlling during crystal growth is still unclear.

In this work, we provide a modified CVT method to easily control the temperature and maintain the growing temperature accurately. A mechanically movable thermal insulation hollow cylinder is stored in the furnace, and its position is auto-tuned by the real-time measured temperatures of the cold end ($T_{\mathrm{cold}}$) for crystal growth. This method provides control of the temperature within ± 0.1 °C throughout the growth process. We successfully grow the single crystals of $MnBi_2Te_4$, $MnBi_4Te_7$, and $MnBi_6Te_{10}$ using this method and search for the differences by finely controlling of the growth temperature. Through magnetic measurements, we find that both the magnetic ground states of $MnBi_4Te_7$ and $MnBi_6Te_{10}$ can be tuned from AFM states to FM states by changing the temperature. The tunable magnetism is caused by the defects in Mn sites. Therefore, a systematic picture of the CVT growth of the MBT family of materials and the magnetic property manipulations has been constructed by our work. In $MnBi_2Te_4$ with optimized magnetism, quantized transport with Chern insulator state is also realized at a low field of 3.7 T. The high reproducibility of the realization of quantized plateau of $h/e^2$ is then confirmed by subsequent transport results in more devices, indicating the high quality of the crystal grown by this modified CVT method. In addition to the MBT family, we also believe this set of CVT growth methods with feedback regulation provides a universally applicable approach to synthesizing more materials with complex crystal phases or sensitivity to growth temperature. The adjustable and varied magnetism in the MBT family makes it possible to explore more exotic quantum effects combining magnetism and topological physics.

## 2. Results and Discussion

By carefully investigating the previous reports of the crystal growth of the MBT

family, we find that the growth temperature is a key parameter for obtaining a specific pure phase of MBT. We then start to aware that more fruitful results on crystal growth may be obtained if the growth temperatures could be controlled more accurately and manipulated more delicately. To achieve the goal of precise control of growth temperature, we modify the CVT growth facility based on a single-zone tube furnace by adding a removable unit and a temperature monitor to the cold end. It needs to be noticed that a two-zone furnace is not chosen because of the small temperature gradient needed for the growth. The construction of the entire homemade crystal growth facility is shown in **Figure 1**a. Specifically, the source materials are sealed and placed on the right side of a quartz ampoule. The ampoule lies in a tube furnace and the right side is aligned to the furnace center. The left end of the ampoule, or called cold end, is placed 110 mm away from the furnace center where the temperature is lower due to the natural temperature gradient of the tube furnace. The cold end of the ampoule is where the single crystal of MBT samples form, and it is inset into a mullite ceramic hollow cylinder, which further blocks the thermal radiation and renders a colder environment inside. The ceramic cylinder is not fixed and can be moved back and forth by a connected stainless rod along with a one-axis translation stage powered by a stepping motor. The changes in temperature following position movement are indicated by the arrows as shown in Figure 1a, i.e., the $T_{cold}$ decreases as the ceramic cylinder moves forward to the hotter region or increases as it moves in the opposite direction. A K-type thermocouple is fixed to the cold end of the ampoule and a nanovoltmeter (Keysight 34420A) collects the real-time temperature readings of the cold end and sends it to a PC. In the PC, a homemade program receives these data and compares them to the temperature setpoint. When the real-time temperature deviates from the setpoint by 0.03 °C, the program communicates with the microcontroller and drives the stepping motor to rotate, thus changing the position of the ceramic cylinder. By this kind of close-loop system, the $T_{cold}$ for crystal growth is detected in real time and can be finely controlled. Figure 1b shows the $T_{cold}$ versus time over 40 hours with and without the feedback regulation as discussed above. One can see that the $T_{cold}$ fluctuates in two oscillated modes, specifically, an oscillation mode with a period of dozens of minutes and another mode with a much longer period of about 12 hours. We believe the former is caused by the temperature control accuracy of the furnace and the latter originated from the ambient temperature difference between day and night. In contrast, when enabling the feedback regulation for temperature control, the temperature fluctuations measured are within ± 0.1 °C, which improves the accuracy of the temperature control by an order of magnitude. Thus, it provides not only a much more stable environment for the high-quality crystal growth of MBT, but also the ability for finely manipulations of the crystal phases and the magnetic properties. According to our early tests, it should be noticed that the attempt to carry out the traditional CVT settings by using a furnace with a temperature control accuracy of 0.1 °C does not really help. Because the cold end is away from the thermocouple equipped at the furnace center, the $T_{\mathrm{cold}}$ would still fluctuate due to the slight change of the ambient conditions and the heat dissipation of the furnace during the weeks of the growth process.

To explore the effects of temperature on the growth of MBT, the starting elements of Mn: Bi: Te in the atomic ratio of 1.7: 2: 4 are mixed with $I_2$ as a transport agent and sealed in a quartz ampoule of 12 mm in inner diameter and 110 mm in length.[42] The ampoule is placed in the furnace with the hot end at the center of the furnace, and the cold end is stretched into the ceramic cylinder for 5 cm in length, and then heated the furnace to the set temperature in 3 h. When the temperature of the hot end is stabilized, the feedback regulation of the $T_{\mathrm{cold}}$ is turned on and keeps working during the entire crystal growth process. After one week of growth, the ampoule is air-quenched and the single crystals are found at the cold end. Controlling the temperature setpoint of the cold end by PC, different phases are obtained in different temperature intervals. For $MnBi_2Te_4$, $MnBi_4Te_7$, and $MnBi_6Te_{10}$, the temperature ranges are ~ 582-592 °C, 569-581 °C and 560-567 °C, respectively. The center of the furnace is set at a temperature 10-13 °C higher than the cold end to maintain the temperature of the hot end as well as to build the rough temperature gradient for vapor transport. The pure phases of the three MBT compounds can be grown stably at the middle region of the corresponding temperature range, and the as-grown samples are in the shape of hexagonal plates or flakes with shiny luster as seen in Figure 1d-f. Mixing of the different phases often occurs at the edges of the different temperature intervals, and $Bi_2Te_3$ in the form of droplets or crystals will form with obvious iodides when the temperature is higher or lower than the growing temperature of MBT. Figure 1c displays the X-ray diffraction (XRD) patterns taken on the cleaved shiny surfaces with (001) lattice planes of the as-grown crystals with the pure phase of $MnBi_2Te_4$, $MnBi_4Te_7$, and $MnBi_6Te_{10}$, where the sharp (00$n$) diffraction peaks can be clearly observed. The lattice constants $c$ calculated from the Bragg equation are ~ 40.9 Å, 23.8 Å, and 101.8 Å respectively, which are consistent with the ones of flux samples and can be used to distinguish the crystal phases. No shift of the diffraction peaks or secondary phases is found, confirming the single-phase nature of our samples.

The synthesis of $MnBi_2Te_4$, $MnBi_4Te_7$, and $MnBi_6Te_{10}$ using the feedback regulation in the CVT method exhibits the effect of fine temperature regulation on single crystalline sample growth of the MBT family. Furthermore, to explore the possibility of tuning the magnetic properties of a single crystal phase, we then grow MBT samples with smaller steps of temperature changes within the $T_{cold}$ range for a corresponding single crystalline phase. For $MnBi_2Te_4$, two subtly different samples are grown at the $T_{cold}$ of 586 °C (sample 1-1) and 581.5 °C (sample 1-2). The temperature-dependent susceptibility $\chi_{H||c}(T)$ is shown in **Figure 2**a, where the magnetic field is applied parallel to the *c* axis with the magnitude of 1 T. The data show the sharp AFM transition takes place at different temperatures of 24.2 K for sample 1-2 and 25.3 K for sample 1-1, suggesting the enhancement of AFM interlayer coupling in the latter compared with traditional flux samples ($T_N \sim 24$ K). This kind of value increase of $T_N$ has also been observed in the previously reported CVT-grown $MnBi_2Te_4$ samples.[43] Figure 2b shows the magnetization of the two batches of crystals of $MnBi_2Te_4$ samples grown under different $T_{cold}$ versus magnetic fields up to 9 T at the temperature of 2 K. The sample 1-2 shows the spin-flop transition at 3.4 T and the magnetic saturation at 7.7 T, while for sample 1-1, the two critical fields are 3.7 T and 8.3 T respectively, which are slightly higher than those for sample 1-2. Also, the magnetization at 9 T for sample 1-1 is 3.77 $\mu_B$/Mn, which is higher than the 3.43 $\mu_B$/Mn for sample 1-2. The increased $T_N$, critical fields, and saturation moments can be related to the defects in samples adjusted by the growth temperature, which would be discussed later, indicating that fine-controlling the growth temperature is a good way to modulate the magnetic properties of the crystals as well as to improve the quality of the crystals.

Although $MnBi_2Te_4$ shows no change of the AFM ground state by varying the growth temperatures, an evolution from the AFM to the FM ground state is observed in $MnBi_4Te_7$. The results are discussed below. Different $MnBi_4Te_7$ samples are grown using $T_{cold}$ below 581.5 °C and we chose three respective samples for discussion (labeled from sample 2-1 to 2-3). The $\chi_{H||c}(T)$ curves in **Figure 3**a-c show the three distinct types of magnetic properties of $MnBi_4Te_7$. For sample 2-1 grown at $T_{cold}$ = 581 °C, the $T_N$ = 13.4 K is found, and a short rise before the overlapping curves diverge at lower temperatures (~ 5 K) is shown in Figure 3a. The $M(H)$ curves exhibit the zigzag shape at 8 K with slight hysteresis, and the hysteresis loop becomes larger with the smaller spin-flip transition from 0.1 T to 0.085 T at 5 K. However, at 3 K, the hysteresis loop is further extended and the spin-flip almost disappears, which means the energy barrier between the AFM state and the FM state is small, and little magnetic fields can overcome the spin-flip transition. At 2 K, a standard hysteresis loop with no magnetic transition from AFM to FM is clearly observed, demonstrating the FM ordering in the $MnBi_4Te_7$. To delineate the two magnetic states with unclear boundaries, we use the temperature corresponding to the minimum value of $d\chi/dt$ as the AFM-FM transition temperature $T_C$. For $T_{cold}$ = 575.6 °C, the $\chi_{H||c}(T)$ of the grown sample 2-2 shows an AFM-FM coexisting state and the enhancement of FM exchange coupling. Compared to sample 2-1, sample 2-2 exhibits lower $T_N$ at 10.5 K and higher $T_C$ at 4.7 K, indicating a trend of transition from AFM to FM states as the decreasing of $T_{cold}$. The $M(H)$ curves are consistent with the $\chi_{H||c}(T)$ curves which suggest AFM transition and the early appearance of FM transition. For sample 2-3 grown at a lower $T_{cold}$ of 568.8 °C, as expected, the $\chi_{H||c}(T)$ presents only FM transition with $T_C$ = 6.8 K. The FM ground state in sample 2-3 is also confirmed by the $M(H)$ data as shown in Figure 3f. According to these results, one can clearly see an evolution of mthe agnetic ground state from AFM to FM in $MnBi_4Te_7$, interrupted by the AFM-FM coexisting state with a change in magnetic transition temperature. The similar AFM to FM magnetic ground state transition is also observed in $MnBi_6Te_{10}$ as the decrease of $T_{cold}$. **Figure 4**a-d show the magnetic properties of $MnBi_6Te_{10}$ grown at $T_{cold}$ = 566.8 °C and 565.5 °C. For sample 3-1 grown at $T_{cold}$ = 566.8 °C, the cusp in $\chi_{H||c}(T)$ suggesting the AFM transition at 11 K. Figure 4c shows that at a temperature below 11 K, kinks appear during the reversal of the magnetic field due to the spin-flip transition, and hysteresis loops appear at lower temperatures. For sample 3-2 grown at $T_{cold}$ = 565.5 °C, the $\chi_{H||c}(T)$ data in Figure 4b shows only FM transition at 5 K which is different from the early reports of the AFM states using self-flux method[24, 27, 44].

As shown above, the magnetic properties in CVT-grown $MnBi_2Te_4$, $MnBi_4Te_7$, and $MnBi_6Te_{10}$ are directly related to the $T_{cold}$. The mechanism for this kind of magnetic property modulation is worth to be identified. According to the reports about the variable magnetic properties in $MnSb_2Te_4$[45-49], Sb-doped $MnBi_4Te_7$[50-52], and $MnBi_6Te_{10}$[40], the magnetic properties are closely related to the Mn content and Mn site defects in the samples.[53] Therefore, it is reasonable to consider that in our experiments, the actual different $T_{cold}$ modulates the Mn content and the defects in the samples, which led to the change of the magnetic properties. We carry out the wavelength-dispersive spectroscopy (WDS) measurements to verify the conjecture that the change in magnetic properties of samples grown at different $T_{cold}$ comes from Mn defects. **Table 1** exhibits the atomic percentages of elements in the samples that have been shown above, suggesting significant differences in Mn content. The sample 1-1, 2-1, and 3-1 with the Mn proportion close to the ideal stoichiometric ratio (14.28% for $MnBi_2Te_4$, 8.3% for $MnBi_4Te_7$, 5.8% for $MnBi_6Te_{10}$) are in the AFM ground state. As the Mn content decreases in the samples grown at lower $T_{cold}$, the AFM coupling weakens and the samples tend to show the FM state, suggesting the correlation between magnetic properties and Mn content in MBT samples, The prevalence of Bi content is greater than the ideal stoichiometric value indicates the presence of Mn/Bi anti-site defects.[34, 47, 54] To clearly show the effect of $T_{cold}$ on the crystal phases and the properties of the magnetism in the MBT family, the magnetic transition temperatures of different phases are summarized in **Figure 5**. The crystal phases are shown on the top and separated by the vertical blue dotted line. The evolution of the AFM transition temperature and FM transition temperature extracted from $\chi_{H||c}(T)$ data cuts the diagram. From $MnBi_2Te_4$ to $MnBi_4Te_7$ and then to $MnBi_6Te_{10}$, the $T_{cold}$ decreases sequentially like the growth temperature used in the self-flux method. In the temperature interval of a single MBT phase, the samples show the $T_{cold}$-dependent magnetic properties due to the presence of Mn defects. For $MnBi_2Te_4$, the differences are reflected in the value of $T_N$, the field for spin flop transition, and the saturation magnetization. The higher the Mn ratio, the larger AFM exchange coupling is found. To confirm that the quality of the sample 1-1 with a higher Mn ratio has indeed improved, few-SLs devices are fabricated by exfoliating from the bulk sample. Figure 5b displays the Chern insulator state that is realized in a two-probe device at 2 K. The plateau of +1 $h/e^2$ appeared at 3.7 T which is lower than the magnetic field required by our former samples grown by flux method.[21] Besides, the high reproducibility of the realization of quantized plateau of $h/e^2$ is also confirmed by subsequent transport results in more devices (Figure S1, Supporting Information). By comparison, the probability of the devices realizing the quantized plateau of $h/e^2$ is normally less than 10% in our previous research when using flux samples. For $MnBi_4Te_7$ and $MnBi_6Te_{10}$, the differences are reflected in the evolution of the magnetic ground state. For both $MnBi_4Te_7$ and $MnBi_6Te_{10}$, the samples contain fewer Mn ions and the magnetic order tends to change from AFM to FM as the decrease of $T_{cold}$. It also needs to be noticed that the elemental ratio of all our results demonstrated in this phase diagram is fixed with Mn: Bi: Te = 1.7: 2: 4, suggesting the great sensitivity of the MBT family to the growth temperature and the potential richer results for further manipulating the degree of freedom on material ratio. To expand the phase diagram and synthesis the other crystal phases such as $MnBi_8Te_{13}$ and $Mn_2Bi_2Te_5$[55-56], the growth parameters such as the temperature gradients and ingredient ratios deserve more studies.

## 3. Conclusion

In conclusion, this work provides a systematical picture for varying the crystal phases and the magnetic properties through growth parameter controlling in the MBT family. Specifically, we synthesize the $MnBi_2Te_4$, $MnBi_4Te_7$, and $MnBi_6Te_{10}$ crystals in the MBT family using a modified CVT method and find the magnetic properties of crystals are sensitive to the cold-end growth temperature. Our results show that the $MnBi_4Te_7$ and $MnBi_6Te_{10}$ can be tunable from AFM to FM ground state by varying the growth temperatures, and the magnetic transition temperature of $MnBi_4Te_7$ tends to be further tuned by fine control of growth conditions. The transition of the magnetic ground state to the FM state is accompanied by the loss of Mn. The $MnBi_2Te_4$ samples grown at different temperatures also show a slight difference in magnetism, and the Chern insulator state is observed in the $MnBi_2Te_4$ device with optimized magnetism at a low magnet field of 3.7 T. In addition, this set of CVT growth method with feedback regulation provides a universally applicable approach that can be applied to the growth of crystals that requires small temperature gradients, high temperature precision, or temperature regulation of a region within a certain temperature background. The tunable and multiple magnetism and topological electronic structure in the MBT family provide excellent material platforms for further topological quantum device fabrication and stimulate more studies on topological physics and investigations of exotic topological quantum states in the near future.

## 4. Experimental Section

Magnetic properties are measured with a Physical Properties Measurement System (Quantum Design) in the temperature range from 2 K to 300 K and magnetic field from -9 T to 9 T using the standard copper sample holders. The magnetic field is applied along the *c* axis of the MBT samples. XRD (Bruker D8 Advance) measurements are performed to identify the crystal phases. To acquire the elemental ratio of samples, the WDS data is measured using Electron Probe Micro Analyzers JXA-8100, and X-ray energy dispersive spectroscopy (EDS) data is collected using an Oxford Inca Energy Dispersive detector equipped on a Quanta 200 Scanning Electron Microscope. The WDS and EDS measurements are performed on the freshly cleaved surfaces of the single crystals using Scotch-tape to avoid any contaminates and oxidation.

## Acknowledgments

The authors gratefully acknowledge the financial support of the National Key Research and Development Program of China (Grant No. 2022YFA1402404); the National Natural Science Foundation of China (Grant Nos. 92161201, T2221003, 12104221, 12104220, 12274208, 12025404, 12004174, 91961101, 61822403, 11874203, 12374043); the Natural Science Foundation of Jiangsu Province (BK20230079); the Fundamental Research Funds for the Central Universities (Grant No. 020414380192).

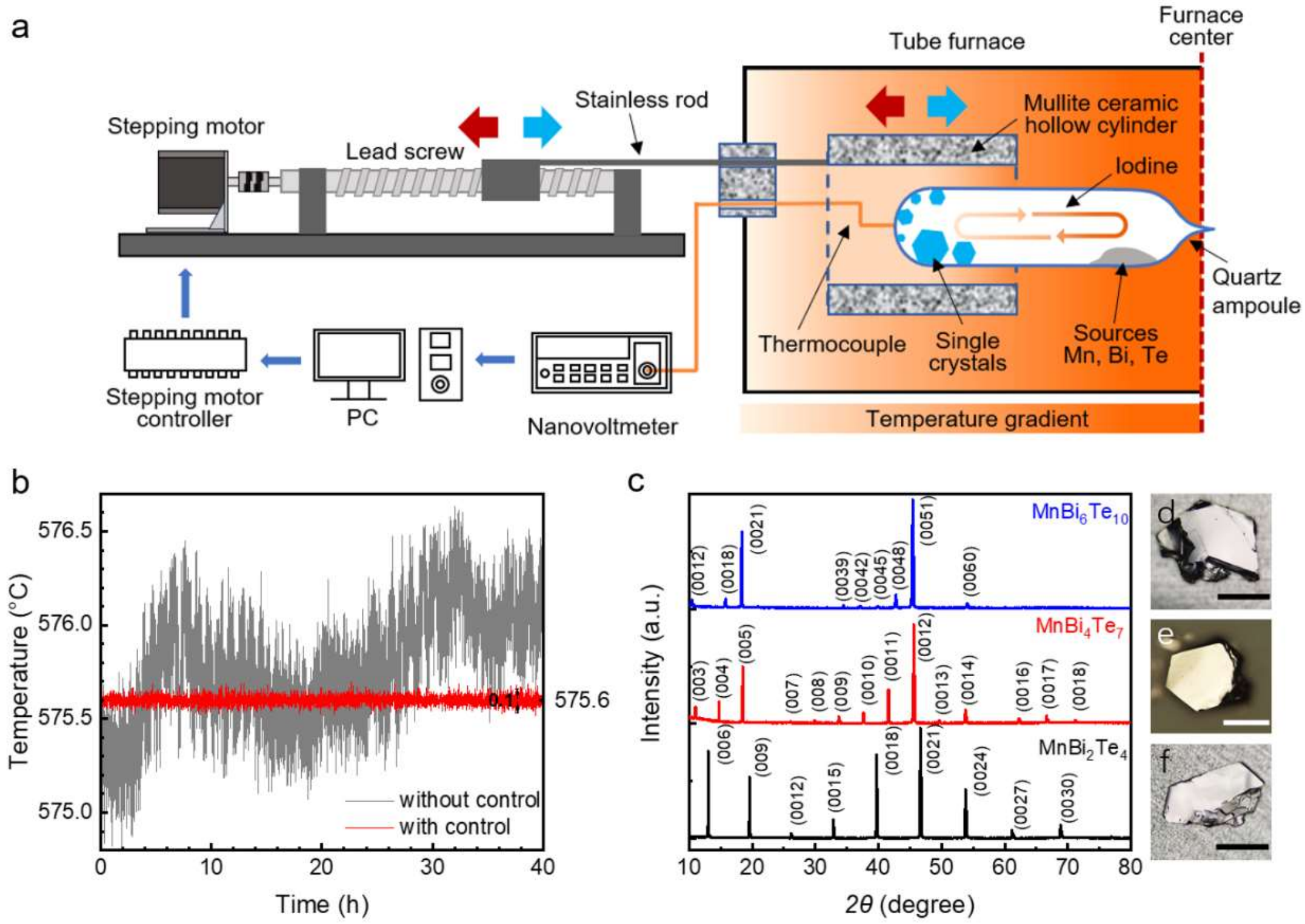

**Figure 1.** Growth method and crystal characterizations. a) Schematic of the growth using the developed tube furnace with feedback regulation. b) Comparison of the temperature fluctuation in the furnace with or without control. c) X-ray diffraction pattern on the (001) surface of different CVT-grown crystal phases. d-f) Optical images of $MnBi_6Te_{10}$, $MnBi_4Te_7$, and $MnBi_2Te_4$ samples, respectively. Scale bar in d-f: 0.4 mm.

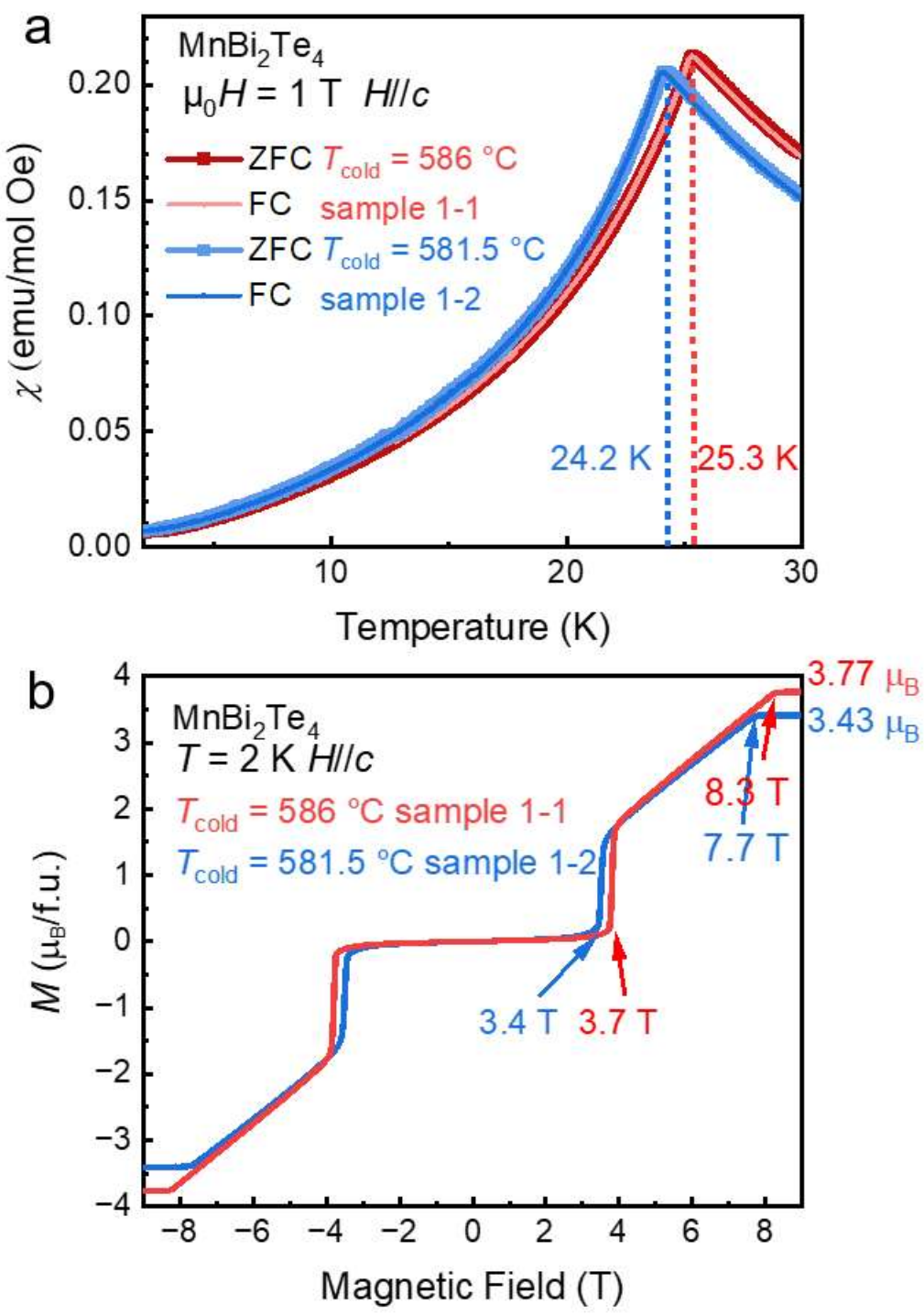

**Figure 2.** The magnetic measurements of $MnBi_2Te_4$ grown in different $T_{cold}$. a) FC-ZFC $\chi_{H||c}(T)$ curves of $MnBi_2Te_4$ grown in the $T_{cold}$ of 586 °C (sample 1-1) and 581.5 °C (sample 1-2) under the magnetic field of 1 T along the *c* axis. b) The field-dependent magnetization $M(H)$ of sample 1-1 and sample 1-1 measured at 2 K along the *c* axis.

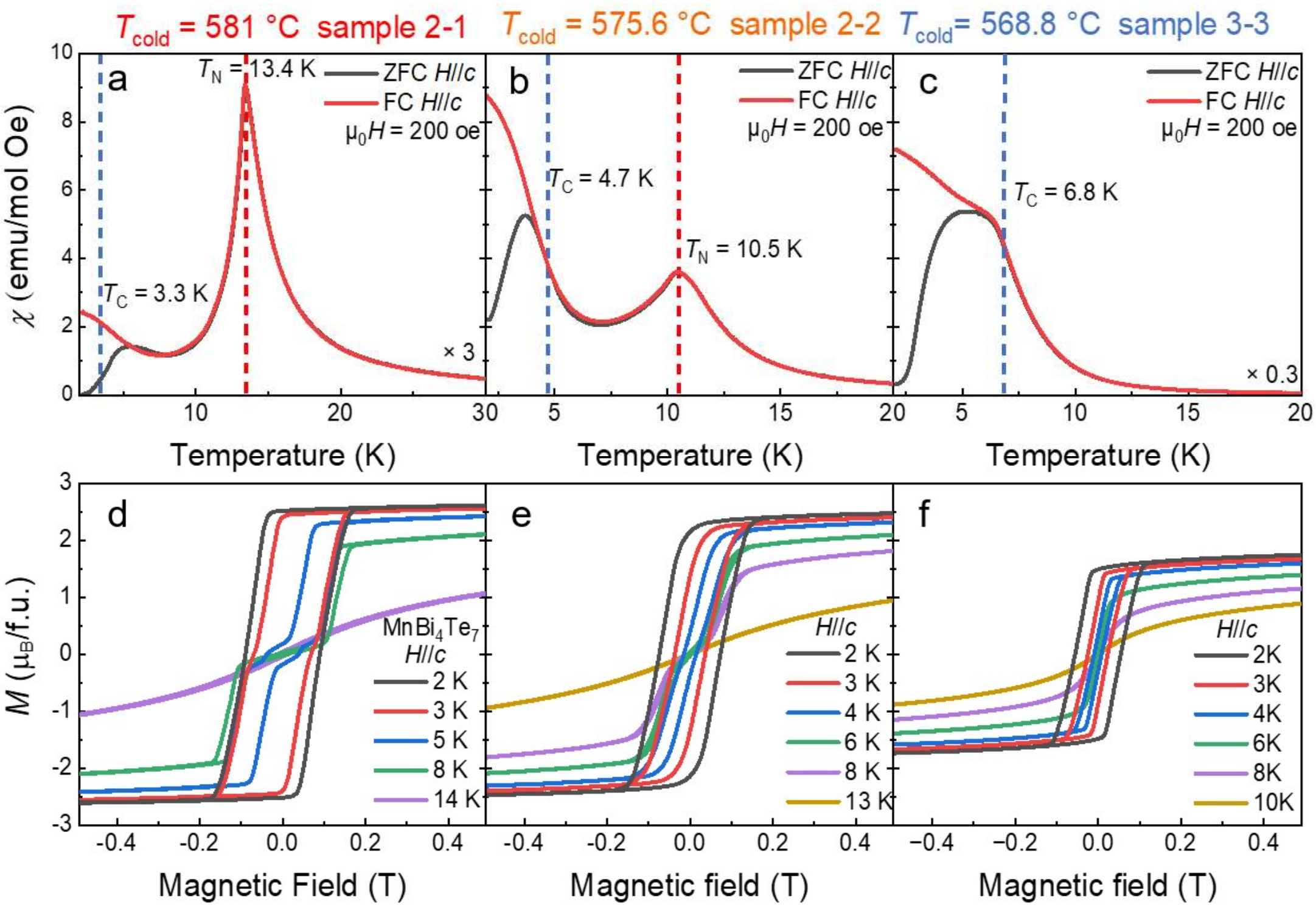

**Figure 3.** The magnetic measurements of $MnBi_4Te_7$ samples grown at various temperatures. a-c) $\chi_{H||c}(T)$ of $MnBi_4Te_7$ grown at $T_{cold}$ of 581 °C (sample 2-1), 575.6 °C (sample 2-2) and 568.8 °C (sample 2-3). d-f) *M*(*H*) of $MnBi_4Te_7$ sample 2-1,2,3 measured at varied temperatures along the *c* axis.

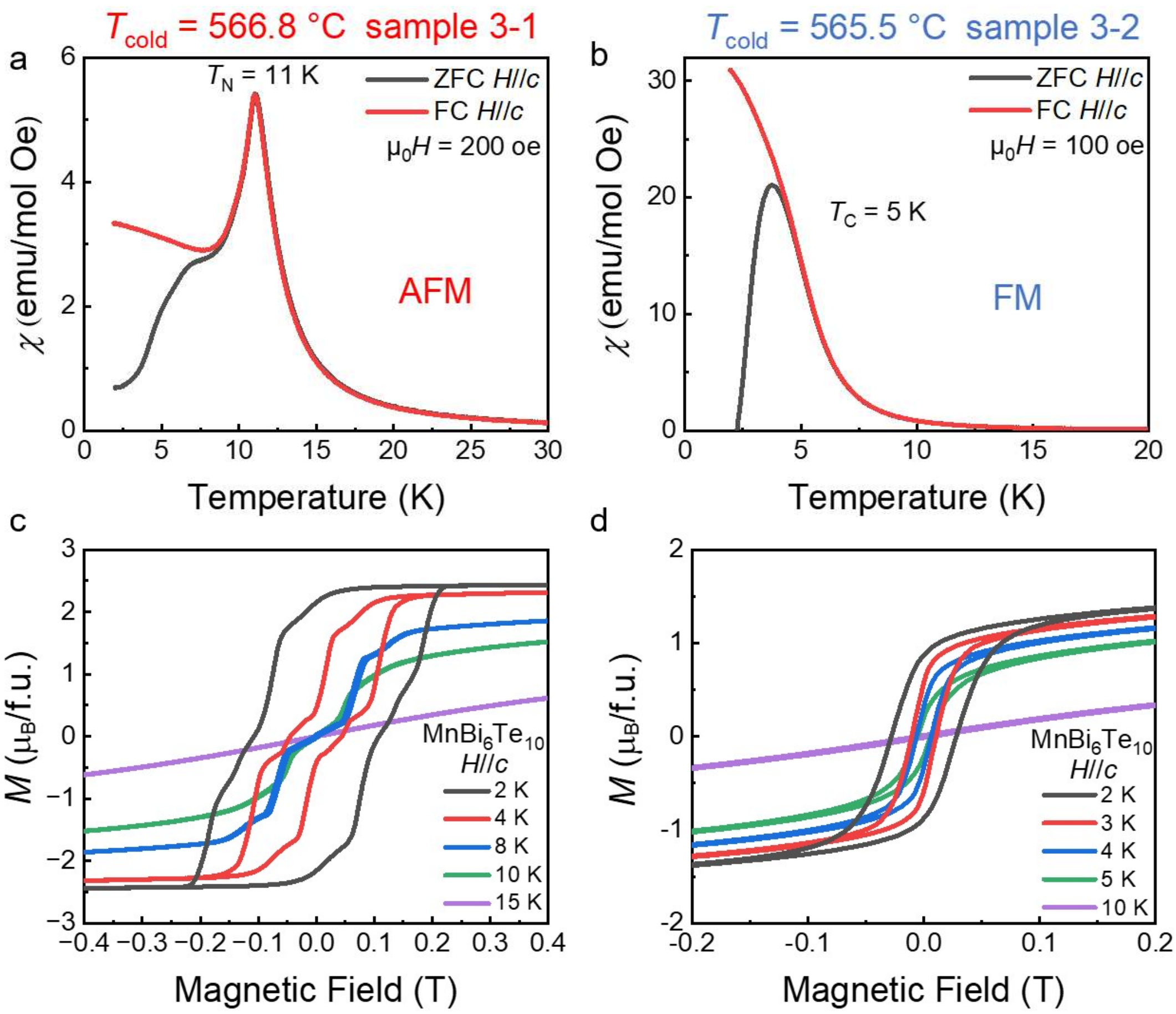

**Figure 4.** The magnetic measurements of $MnBi_6Te_{10}$ grown in different $T_{cold}$. a,b) $\chi_{H||c}(T)$ of $MnBi_2Te_4$ grown in the $T_{cold}$ of 566.8 °C (sample 3-1) and 565.5 °C (sample 3-2). c,d) $M(H)$ of sample 3-1,2 measured at varied temperatures along the $c$ axis.

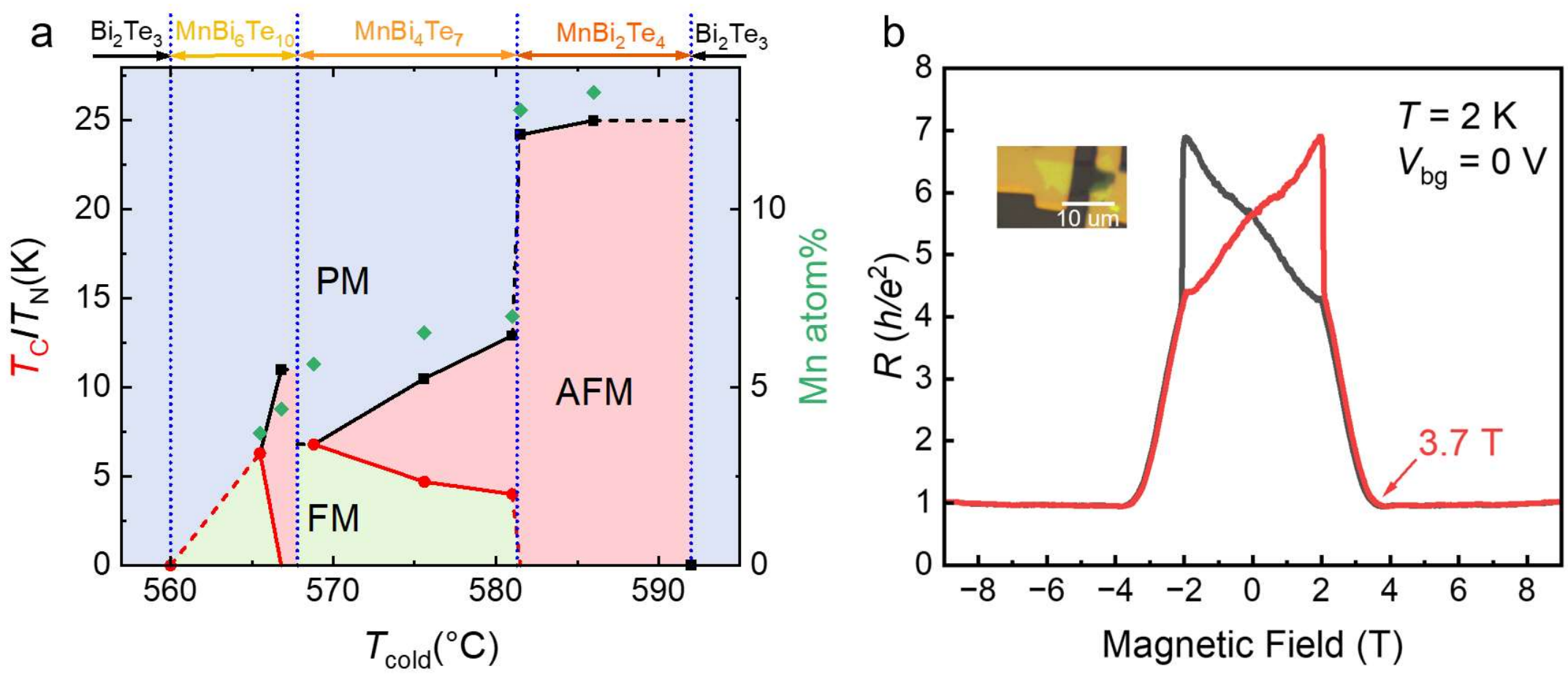

Figure 5. The crystal growing phase diagram and Chern insulator state in the CVT grown $MnBi_2Te_4$. a) The crystal growing phase diagram with the magnetic phase transition and the corresponding Mn atom% from WDS. b) The two-probe resistance R measured at 2 K and $V_{bg}$ = 0 V, with a plateau of +1 $h/e^2$ appeared at 3.7 T. The inset is the optical image of the device.

**Table 1.** WDS results of MBT samples. The elemental ratios are written as the atom% from WDS measurements. The cold-end temperature $T_{cold}$, Néel temperature $T_N$, and FM transition temperature $T_C$ are attached to discriminate between different kinds of samples.

| Sample number | Crystal Phase | $T_{cold}$ (°C) | $T_N$ (K) | $T_C$ (K) | Mn atom% | Bi atom% | Te atom% |
|---|---|---|---|---|---|---|---|
| 1-1 | $MnBi_2Te_4$ | 586 | 25.3 | -- | 13.29 | 30.47 | 56.24 |
| 1-2 | $MnBi_2Te_4$ | 581.5 | 24.2 | -- | 12.79 | 30.51 | 56.70 |
| 2-1 | $MnBi_4Te_7$ | 581 | 13.4 | 3.3 | 7.00 | 34.93 | 58.07 |
| 2-2 | $MnBi_4Te_7$ | 575.6 | 10.5 | 4.7 | 6.54 | 35.88 | 57.58 |
| 2-3 | $MnBi_4Te_7$ | 568.8 | -- | 6.8 | 5.65 | 37.29 | 57.06 |
| 3-1 | $MnBi_6Te_{10}$ | 566.8 | 11 | -- | 4.39 | 37.98 | 57.63 |
| 3-2 | $MnBi_6Te_{10}$ | 565.5 | -- | 5 | 3.72 | 38.96 | 57.32 |